\documentclass[prd,
nofootinbib,
preprintnumbers,
twocolumn,
showpacs
]{revtex4}
\usepackage{graphicx,epsf,epsfig}
\def\eq#1{{eq.~(\ref{#1})}}

\def\hbar{\hspace{0pt}\raisebox{1pt}{$-$} \hspace{-7pt} h}

\def\5{\overline 5}

\newcommand{\be}{\begin{equation}}
\newcommand{\ee}{\end{equation}}
\newcommand{\bd}{\begin{displaymath}}
\newcommand{\ed}{\end{displaymath}}
\newcommand{\bea}{\begin{eqnarray}}
\newcommand{\eea}{\end{eqnarray}}
\newcommand{\nn}{\nonumber}

\begin{document}
\title{On CP Violation in Minimal Renormalizable SUSY SO(10) and Beyond}
\date{April 26, 2005}
\author{Stefano Bertolini }\email{bertolin@he.sissa.it}
\author{Michal Malinsk\'{y}}\email{malinsky@sissa.it}
\affiliation{Scuola Internazionale Superiore di Studi Avanzati,
via Beirut 4, I-34014 Trieste and INFN, Sezione di Trieste, Italy}
\begin{abstract}
We investigate the role of CP phases within the renormalizable
SUSY $SO(10)$ GUT with one $10_{H}$, one $\overline{126}_{H}$ one
$126_{H}$ and one ${210}_{H}$ Higgs representations and type II
seesaw dominating the neutrino mass matrix. This framework is non
trivially predictive in the fermionic sector and connects in a natural way the GUT
unification of  $b$ and $\tau$ Yukawa couplings with the bi-large
mixing scenario for neutrinos. On the other hand, existing numerical
analysis claim that consistency with quark and charged lepton data prevents the minimal setup
from reproducing the observed CP violation via the Cabibbo-Kobayashi-Maskawa (CKM) matrix.
We re-examine the issue and find by inspection of the fermion mass sum rules
and a detailed numerical scan that, even though the CKM phase preferentially takes
values in the second quadrant, the agreement of the minimal model
with the data is actually obtained in a non negligible fraction of
the parameter space. We then consider a recently proposed renormalizable
extension of the minimal model, obtained by adding one chiral
120-dimensional Higgs supermultiplet. We show that within such a setup the CKM phase falls
naturally in the observed range. We emphazise the robust predictivity
of both models here considered for neutrino parameters that are in the reach
of ongoing and future experiments.

 \end{abstract}
\maketitle

\section{Introduction}
Testing neutrino properties is one of the greatest challenges of the contemporary high
energy physics.
Thanks to the complementary informations provided by solar, atmospheric and reactor
based neutrino experiments, we have obtained in the last few years a consistent description of
the neutrino phenomenology in terms of the phenomenon of oscillations.
The fact that the observed neutrino mass and mixing pattern differs drastically from those
typical of the quark sector is remarkable and at the same time intriguing.
On the one hand, the typical scale of the neutrino masses (more precisely
of neutrino mass differences) lies below the electronvolt scale. On the other hand, large (almost
maximal) mixings appear in the lepton sector, at variance with the hierarchical structure
of the quark sector.
The precise data that we have obtained in the last years on neutrino properties
has renewed and spurred the interest on the basic question about the origin
of fermion masses and mixings, that remains unanswered within the Standard Model (SM) of electroweak
interactions.

The smallness of the neutrino mass may be naturally related to a large Majorana mass scale for
the right-handed (RH) components via the seesaw mechanism~\cite{general-seesaw-type-I}.
This mechanism is naturally
embedded in grand unified scenarios. Of particular interest are Grand Unified Theories (GUT)
based on $SO(10)$, where all the known fermions, including the RH neutrino components,
are contained in three 16-dimensional fundamental spinorial representations. In such a framework
the RH Majorana mass is identified with the left-right symmetry breaking scale.

In the last few years a supersymmetric (SUSY) GUT model based on the $SO(10)$ gauge group has
attracted renewed interest for its sharp predictivity of neutrino
observables~\cite{Aulakh:2003kg,Bajc:2004xe}.
The model has no more parameters than the Minimal Supersymmetric Standard Model (MSSM) with massive
neutrinos and exact R-parity.
In the minimal renormalizable setting the model contains three generations
of $16_F$ matter supermultiplets and the following Higgs chiral
supermultiplets~ \cite{Clark:ai,Aulakh:1982sw}: $10_H$, $\overline{126}_H$, $126_H$, and $210_H$.
The $10_H$ and $\overline{126}_H$ representations
couple to the matter bilinear $16_{Fi} 16_{Fj} = (10_S + 120_A +{126}_S)_{ij}$ in the superpotential leading to
the minimal set of Yukawa couplings needed for a
realistic fermion mass spectrum~\cite{Babu:1992ia}
($S,A$ denote the symmetry property in the generation indices $i,j$).
The $\overline{126}_H$ multiplet also contains a left-handed (LH) Higgs triplet that induces small
Majorana neutrino masses via type II seesaw~\cite{general-seesaw-type-II}.
The $126_H$ representation
is needed in order to preserve supersymmetry from $D$-term breaking, while
the $210_{H}$ triggers the
$SO(10)$ gauge symmetry breaking and provides the needed mixings
among the Higgs supermultiplets.

An attractive property of the model is that when dominance of type II seesaw is invoked~\cite{Bajc:2004fj}
the maximality of the atmospheric neutrino mixing
can be linked to the $b-\tau$ Yukawa coupling unification \cite{Bajc:2001fe,Bajc:2002iw}.
The model features exact R-parity conservation, due
to the even congruency class ($B-L=2$) of the $10$ and  $\overline{126}$
representations (the $120$ representation shares the same property),
with relevant implications for cosmology and proton
decay~\cite{Lee:1994je,Aulakh:1997fq,Aulakh:1999cd,Aulakh:2000sn,Nath:2001uw,Bajc:2003ht,Goh:2003nv,Fukuyama:2004pb}.

In a previous paper~\cite{Bertolini:2004eq} we studied in detail the implications of the mass sum rules of the
model on the neutrino mass parameters. We showed that in the case of CP conserving
Yukawa couplings (advocating for the sake of the discussion the soft supersymmetry breaking sector
as the source of the observed CP violation), when 2-$\sigma$ ranges for the quark masses and mixings are considered, the model
is consistent with all lepton data at the level of the present accuracy.
On the other hand, when the analysis is limited to
1-$\sigma$ ranges the model shows clear tensions in reproducing
the observed leptonic spectrum. In particular, the electron mass is reproduced by an extreme fine tuning among quark parameters,
while the solar mixing angle $\theta_{12}$ is predicted too close to
maximal, and the deviation of the atmospheric mixing $\theta_{23}$ from maximal is
too large. The shortcomings of the model were previously emphasized in the
literature~\cite{Matsuda:2000zp,Mohapatra1}, albeit considering a too limited
range of uncertainties in the input data.

In the same paper we worked out a simple renormalizable extension of the
SUSY $SO(10)$ model by including
one $120_H$ supermultiplet to complete the allowed Yukawa interactions. On the basis of analytic arguments
and numerical analysis
we showed that a small $120_H$ contribution to the Yukawa potential allows for a substantial improvement
on the fit of the lepton mixings even at the 1-$\sigma$ level.
At the same time, the set of the model parameters remains overconstrained by the input data, and
the framework provides non-trivial outcomes in the neutrino sector.

When complex Yukawa couplings are taken into account, recent
analyses~\cite{Mohapatra2,Mohapatra3,Dutta:2004hp} show
that the fit of the lepton mixings may improve, but
the Cabibbo-Kobayashi-Maskawa (CKM) phase is forced to the second or third
quadrant by the electron mass fit,
thus requiring significant contributions to CP violation from other sources.
Non-renormalizable operators are invoked in ref.~\cite{Mohapatra3},
while in ref.~\cite{Dutta:2004hp}
a $120$ dimensional Higgs extension is considered (with an additional parity symmetry),
in order to restore the agreement with the data.
CP violation in a similar SUSY $SO(10)$ framework with a $U(2)$ family symmetry was discussed
in ref.~\cite{Chen:2001pr}.

In this paper we reconsider the study of the fermion mass and mixing patterns in the minimal
renormalizable SUSY $SO(10)$ model with complex Yukawa couplings.
We comment on the simple analytic argument that justifies the CKM phase taking preferentially values
in the second or third quadrant and show,
by a careful numerical treatment of the fermion mass rules, that the minimal model exhibit
no dramatic tension among quark, charged leptons data and CKM CP violation. The fit of neutrino data
shows some tension with the value of the strange quark mass which is required to be large in order
to reproduce the recently increased lower limit on the solar neutrino mixing.
Nevertheless the minimal framework does remain a viable GUT candidate\footnote{Due to the
large chiral super multiplets present in the model the running gauge coupling diverges shortly above the
GUT scale. This may call for additional unknown physics to enter below the Planck scale \cite{Chang:2004pb} or for an
effective gravitational scale to enter near the GUT scale. Such a discussion is beyond the scope of this paper
in as much as perturbativity is maintained up to the GUT scale.}.
In particular, we emphasize the sharpness and
the robustness of the prediction of the $U_{e3}$ lepton
mixing, that is bound to be non-vanishing and within the reach of planned neutrino experiments.

We conclude the discussion by considering the renormalizable $120_H$ extension of the minimal
setting proposed in ref.~\cite{Bertolini:2004eq}. While the authors of ref.~\cite{Dutta:2004hp},
in order to sensibly reduce the number of parameters,
impose an additional parity symmetry that makes all Yukawa couplings hermitian,
in ref.~\cite{Bertolini:2004eq} it is assumed that the $120_H$ induced contributions
to fermion masses are a perturbation of the minimal scenario (from percent to 0.1 percent level).
It was shown that the additional (antisymmetric) Yukawa interaction leads to a dramatic
improvement on the fit of the fermion masses and mixings,
while maintaining, as a perturbation, the predictivity of the minimal framework.
Here we address CKM CP violation and show that in the extended model the agreement with the data is easily obtained,
while maintaining a robust non vanishing lower bound for the $U_{e3}$ entry of the lepton mixing.

\section{Fermion masses and mixing in the minimal renormalizable model}
Henceforth, unless otherwise stated, we will follow and refer to the notation of ref.~\cite{Bertolini:2004eq}.
When type II seesaw is considered, the mass matrices for the SM fermions in the
minimal renormalizable SUSY $SO(10)$ scenario are given by~\cite{Babu:1992ia,Mohapatra:1979nn}
\bea
M_u &=& Y_{10}v_u^{10}+ Y_{126}v_u^{126}\nn ~,\\
M_d &=& Y_{10}v_d^{10}+ Y_{126}v_d^{126}\label{minimalrelations} ~,\\
M_l &=& Y_{10}v_d^{10}-3 Y_{126}v_d^{126}\nn ~,\\
M_\nu &=& Y_{126}v_T^{126} ~,\nn
\eea
where $Y_{10}$ and $Y_{126}$ are complex symmetric 3x3 matrices, $v_{u,d}^{10,126}$ denote the VEVs
of the components of the $10_{H}$ and $\overline{126}_{H}$ multiplets that enter the light Higgs
doublets of the MSSM, and $v_T^{126}$ is the tiny induced VEV of the LH triplet component in $126_H$.
These relations can be translated into the following GUT scale sum rules for the charged lepton and (Majorana)
neutrino mass matrices~\cite{Mohapatra2}:
\bea \label{sumrule1}
k \tilde{M_{l}}=\tilde{M}_{u}+r \tilde{M}_{d} \qquad M_{\nu}\propto M_{l}-M_{d}
\eea
where $k$ and $r$ are in general complex ${\cal O}(1)$ functions of $v_{u,d}^{10,126}$, and the tilded
matrices are normalized to their maximal eigenvalue.
Let us first rescale by a global phase eq. (\ref{sumrule1}) such
that $r$ becomes real ($r=-|r|{\rm e}^{i\phi_{r}}$) and define $k'= k\, {\rm e}^{-i \phi_{r}}$.
Since all the mass matrices are
symmetric they can be diagonalized by means of unitary transformations $M_{x}=U_{x}^{T} D_{x} U_{x}$.
Taking into account that
$U_{u}^{T}U_{d}^{*}\equiv V_{CKM}^{0}=P_{u}V_{CKM}P_{d}$, where
$P_{u}={\rm diag}({\rm e}^{\frac{1}{2} i\alpha_{1}},{\rm e}^{\frac{1}{2} i\alpha_{2}},{\rm
e}^{\frac{1}{2} i\alpha_{3}})$ and $
P_{d}={\rm diag}({\rm e}^{\frac{1}{2} i\beta_{1}},{\rm e}^{\frac{1}{2} i \beta_{2}},1)$
parametrize the
re-phasing of the quark mass matrices necessary to bring the CKM matrix in the standard form (denoted by
$V_{CKM}$ with one CP-violating phase, $\delta_{CKM}$) and lead to positive eigenvalues,
we may rewrite the sum-rules in the $M_{d}$-diagonal basis as follows
\bea \label{sumrulesindbasis}
k' \tilde{M_{l}'}&=&{V_{CKM}^{0}}^{T}\tilde{D}_{u}V_{CKM}^{0}-|r| \tilde{D}_{d}  \\
k' \tilde{M}'_{\nu} &\propto&
{V_{CKM}^{0}}^{T}\tilde{D}_{u}V_{CKM}^{0}-
|r|\tilde{D}_{d}
- {\rm e}^{i\omega}\left|k' \frac{m_{b}}{m_{\tau}}\right|
\tilde{D}_{d} \nn
\eea
The primed matrices in eq. (\ref{sumrulesindbasis}) are given by $M_{x}'\equiv U_{d}^{*}M_{x}U_{d}^{\dagger}$,
while $\omega$ accounts for the phase of $k'$ and the sign of $m_{b}/m_{\tau}$.
The factors $\frac{1}{2}$ in definitions of phases of $P_{u,d}$ are chosen to maintain compatibility with the
notation used in ref.~\cite{Mohapatra2}.

One can now exploit the information contained in eq. (\ref{sumrulesindbasis}) by plugging into the
charged lepton mass formula all
the parameters that are known (by running the data up to the GUT scale), namely the ratios
of quark masses in $\tilde{D}_{u}\equiv {\rm diag}(|m_{u}/m_{t}|,|m_{c}/m_{t}|,1)$,
 $\tilde{D}_{d}\equiv {\rm diag}(|m_{d}/m_{b}|,|m_{s}/m_{b}|,1)$, the CKM mixing angles and the
CP-violating phase $\delta_{CKM}$
and vary them within their experimental ranges. The remaining 8 real parameters $|r|$, $|k'|$,
$Arg(k')$, $\alpha_{i}$, $\beta_{j}$, that appear in the first mass sum-rule,
in principle arbitrary are varied over their allowed domains.
In spite of the many parameters, the charged lepton mass sum rule (\ref{sumrulesindbasis}) is overconstrained and
some fine tuning is needed to reproduce the lighter eigenvalues.
Having obtained an allowed solution of the charged lepton mass sum rule,
the neutrino masses and mixings are then sensitive to
the sign of $m_{b}/m_{\tau}$ and the phase of $k'$ ($\omega$). Since a negative interference
between the 3-3 entries of the $M_l$ and $M_d$ matrices is needed to obtain a large atmospheric
mixing angle, the phase $\omega$ becomes strongly correlated to quark phases.
As a consequence this framework is highly constrained in the neutrino sector as well and determines characteristic
correlations among the neutrino parameters.

In the light of the above remarks it is undoubtedly intriguing that
this scenario is shown to fit the main features of the SM data
even when Yukawa phases are neglected~\cite{Mohapatra1,Bertolini:2004eq}.
As we have already mentioned,
in such a case, when considering 1-$\sigma$ uncertainties in the input parameters,
a rather stringent lower bound for the solar
mixing angle $\sin^{2}2\theta_{12}\gtrsim 0.85$ appears
and the atmospheric mixing angle can hardly be maximal.
In addition one finds a lower bound for the $|U_{e3}|$
parameter $|U_{e3}|\gtrsim 0.15$, just at the value of the
present 90\% C.L. experimental upper bound $|U_{e3}|\lesssim  0.15$.
While the constraints on the solar and atmospheric mixings are substantially relaxed when
considering 2-$\sigma$ uncertainties in the input data, the lower bound
on $|U_{e3}|$ is robust and represents a characteristic feature of the model
as discussed in \cite{Mohapatra1,Bertolini:2004eq}.

Before proceeding to the discussion of numerical results let us review some issues
related to the fit of the fermion spectrum in the case of complex Yukawa couplings.

\subsection{$\delta_{CKM}$ and the electron mass formula}

When complex Yukawa couplings are considered, it is claimed that a
successful fit of the electron mass forces the CKM phase to take values in the second or third
quadrant~\cite{Mohapatra2,Mohapatra3,Dutta:2004hp},
thus requiring an extension of the model to recover the measured CP violation.

The argument that supports the numerical outcome is based on the approximate
formula for the electron mass eigenvalue that can be obtained from
(\ref{sumrulesindbasis}) using the hierarchical properties of the quark mass matrices in
the right-hand side (RHS):
\bea
|k'\,\tilde m_{e}|{\rm e}^{i \phi_{1}}=
-|r| \,{\rm e}^{i \beta_{1}}F_{d}\lambda^{4}+{\rm e}^{i \alpha_{2}} F_{c}\lambda^{6}\nn\\
-A^{2}\Lambda^{2}{\rm e}^{i \alpha_{3}}\lambda^{6}
\frac{|r|}{{\rm e}^{i \alpha_{3}}-|r|}+{\cal O}(\lambda^{7})
\label{meminimal}
\eea
Here $\Lambda\equiv(1-\rho-i\eta)$ where $\rho$ and $\eta$ are the Wolfenstein CKM parameters,
while $F_{d}\equiv \frac{m_{d}}{m_{b}}/\lambda^{4}$, $F_{c}\equiv \frac{m_{c}}{m_{t}}/\lambda^{4}$
are ${\cal O}(1)$ factors.
Fitting the normalized electron mass (with a typical magnitude of the order of
 ${\cal O}(\lambda^{5})$) amounts to compensating
the dominant $\lambda^{4}$ term in the RHS by other terms therein, the only possible one
at the given order of expansion being
that proportional to $\Lambda^{2}$. In turn this amounts to constraining the size of the CKM phase.
The CKM phase is encoded in $\rho$ and $\eta$ as
${\rm e}^{i\delta_{CKM}}s_{13}=A\lambda^{3}(\rho+i\eta)$.
The typical values of $\rho$
and $\eta$ for $\delta_{CKM}$ in the physical region are centered around $\rho\sim 0.21$ and
$\eta\sim 0.34$. Since the parameter $|\Lambda|^{2}={(1-\rho)^{2}+\eta^{2}}$
is maximized for $\rho < 0$
(in refs. \cite{Mohapatra1,Bertolini:2004eq} the CP conserving case $\eta=0$ was considered),
the fit of the electron mass in formula (\ref{meminimal}) seems to strongly disfavor
the CKM phase in the first quadrant \cite{Mohapatra2}.

On the other hand, one should be careful in claiming the relevance of
``subleading'' terms in a truncated expansion.
A detailed inspection of the $O(\lambda^7)$ terms in \eq{meminimal}
shows that $\lambda$ is not a faithful
expansion parameter, in that some cofactors, not necessarily dependent on $\Lambda$, can become
accidentally large (a small denominator in the $\lambda^6$ term is an example).
Therefore a larger number of ``subleading'' terms in \eq{meminimal} may contribute on top of the
${\cal}O(\lambda^{6})$ term, and the scan over the complex phases must be very
detailed not to miss such a solutions
(we stress that in the minimal model the electron mass is nevertheless the outcome of a fine tuning; what we
are here discussing is the extent).
As an example let us consider the ${\cal O}(\lambda^{7})$ term
\be
{\cal O}(\lambda^{7})\sim
\frac{A^{4} \Lambda^{2}\ |r|\ e^{i(2\alpha_{3}-\beta_{2})}}{F_{s}(e^{i\alpha_{3}}-|r|)^{2}}\lambda^{7}
+\ldots \ ,
\label{lambda7}
\ee
where $F_{s}\equiv \frac{m_{s}}{m_{b}}/\lambda^{3}$.
Since for typical values of $r$ one obtains $(1-|r|)^{2}\approx \lambda^2$,
the denominator can be small enough to lead to an important correction to the ${\cal
O}(\lambda^{6})$ term. Notice that a small $m_s$ favors as well the needed destructive
interference in \eq{meminimal}, as emphasized in ref.~\cite{Mohapatra3}.

In our numerical analysis we pay particular attention to the quality of the
parameter scan in those regions that may lead to
departures from the expectations based on the size of $\Lambda$ in \eq{meminimal}.
Once an approximate solution of the charged lepton mass sum-rule is found (at the few percent level)
a detailed analysis is performed in the neighbor parameter space by linearizing the mass relations.
Such a procedure improves dramatically the convergence of the numerical code, revealing
solutions that would escape the original scan, unless one performs it with extremely high
granularity and huge demand of computing power.
Indeed, such an improved numerical analysis shows many solutions of the
charged lepton mass formula emerging in
parts of the parameter space where the cancelation among the leading terms in eq. (\ref{meminimal})
is not as effective. We discuss and show our numerical results in the next subsection.

The sensitivity of the electron mass fit to subleading terms in eq. (\ref{meminimal})
is crucial in the case of the extended model considered in ref.~\cite{Bertolini:2004eq}
with an additional 120-dimensional chiral super-multiplet.
As we shall see in Section \ref{modelwith120}, even for typical
magnitudes of the $120_H$ Yukawa contributions to fermion masses as low as 0.1\% of those coming from the 10 and 126
dimensional Higgs multiplets, the role of the $\Lambda^{2}$ term in eq.~(\ref{meminimal})
is easily screened by the $120_H$ induced terms, thus lifting the bias on the CKM phase.

\subsection{Numerical results}
In the first stage we performed a detailed fit of the
relations (\ref{sumrulesindbasis}).
We use GUT scale input data for quark and leptons as derived in ref.~\cite{Das:2000uk}
via two-loop renormalization group analysis.
Table~\ref{GUTdata} shows a sample of quark and charged lepton parameters and their uncertainties,
for a given choice of the supersymmetric threshold scale $M_S$ and $\tan\beta$. A supersymmetric
threshold scale $M_S\simeq 1$ TeV and $\tan\beta=10,\ 55$ will be considered as typical values
for the MSSM gauge coupling unification underlying the one-step $SO(10)$ breaking scenario
here considered. The GUT relations between fermion masses and Yukawa couplings depends on the
vacuum expectation values (VEV). Henceforth we follow the $\overline{\rm MS}$ prescription adopted
in ref.~\cite{Das:2000uk}, to which we refer the reader for details.

\begin{table}
\caption{\label{GUTdata}
Sample of GUT-scale values for quark masses and mixings and charged lepton masses. 
The 1-$\sigma$ data are shown for a GUT scale $M_G = 2\times 10^{16}$ GeV, a supersymmetric
threshold scale $M_S = 1$ TeV and $\tan\beta=10$~\cite{Das:2000uk,Fusaoka:1998vc,PDB}.
We use the results of ref.~\cite{Das:2000uk} up to the $u$, $d$ and $s$ quark masses
whose central values and uncertainties are updated according to the discussion in the text.
In our analysis we scan over $90\%$ C.L. ranges ($1.64\ \sigma$) of the input parameters. 
}
\begin{tabular}{@{\hspace{10pt}}c@{\hspace{20pt}}r@{\hspace{10pt}}}
\hline \hline
$m_{e}$ & 0.3585  MeV \\
$m_{\mu}$ & $75.67^{+0.06}_{-0.05}$  MeV \\
$m_{\tau}$ & $1292.2^{+1.3}_{-1.2}$  MeV \\
\hline
$m_{d}$ & $1.5^{+0.42}_{-0.23}$ MeV \\
$m_{s}$ & $29.94^{+4.30}_{-4.54}$  MeV \\
$m_{b}$ & $1.06^{+0.14}_{-0.08}$  GeV \\
\hline
$m_{u}$ & $0.72^{+0.13}_{-0.14}$  MeV \\
$m_{c}$ & $210.32^{+19.00}_{-21.22}$  MeV \\
$m_{t}$ & $82.43^{+30.26}_{-14.76}$  GeV \\
\hline
$\sin \phi_{12}$ & $0.2243\pm 0.0016$  \\
$\sin \phi_{23}$ & $0.0351\pm 0.0013$  \\
$\sin \phi_{13}$ & $0.0032\pm 0.0005$  \\
$\delta_{CKM}$ & $60^{o}\pm 14^{o}$ \\[0.5ex]
\hline
\hline
\end{tabular}
\end{table}

We consider $90\%$ C.L. ranges for all input parameters. With respect to the 
data used in refs.~\cite{Das:2000uk,Fusaoka:1998vc} we account for 
larger uncertainties in the masses of the light $u,\ d$ quarks and update the range
of $m_s$. While $m_u$ plays a subleading role in the mass sum rules, a light
$m_d$ favors the reproduction of the electron mass eigenvalue, as we discussed
in the previous subsection. The values of $m_u$ and $m_d$ given in Table I
correspond to $4.88\pm 0.57$ MeV and $9.81\pm 0.65$ MeV respectively at $1$ GeV~\cite{Fusaoka:1998vc}.
The present ranges for the $\overline{\rm MS}$ running masses at $1$ GeV
of the up and down quarks are $2$ to $5.4$ MeV and  $5.4$ to $10.8$ MeV respectively~\cite{PDB}. 
Accordingly, we will allow for values of $m_d$ at the
GUT scale as low as $0.7$ MeV. As for the strange quark mass an up to date range,
which includes the lattice evaluations, is $m_s(2\ {\rm GeV}) = 110\pm 20$ MeV, corresponding 
at the GUT scale to $23\pm 6$ MeV.

\begin{figure}[b]
\caption{\label{figure_density} The relative density spectrum of allowed
solutions for the charged lepton masses is shown as a function of the
CKM phase $\delta$.
Although there appear a preference for the quadrants with $\rho < 0$,
there exists a significant number of solution with $\delta_{CKM}$
in the physical region.
Quantitatively the ratio of the solution density in the preferred area
(the darkest slice) to that in the 1-$\sigma$ range is about 7 to 1.}
\epsfig{file=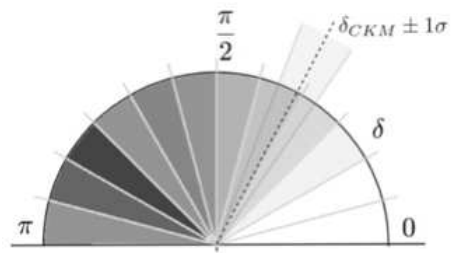 ,width=6cm}
\end{figure}

The complex phases $\alpha_{2,3}$,
$\beta_{1,2}$
are sampled in the whole range $[0,2\pi)$. The phase $\omega$ shows then a correlation to the
quark phases as a consequence of the tight relation between large
atmospheric and $b-\tau$ Yukawa unification, which involves a partial cancelation
among the terms in the RHS of the neutrino mass sum rule.
Since the reduced up-quark mass
$\tilde{m}_{u}$ is by far the smallest parameter in the mass relations, $\alpha_{1}$
does not play any relevant role.

In Fig. \ref{figure_density} we display the density spectrum of the solutions of the charged
lepton mass formula in eq. (\ref{meminimal}) as a function of
$\delta_{CKM}$ in the interval $[0,\pi)$.

As one sees the relative density of solutions of the charged
lepton sum rule is far from being negligible even when considering $\delta_{CKM}$
in the 1-$\sigma$ range, contrary to the claims in refs.
\cite{Mohapatra2,Mohapatra3}. To make this statement quantitative
the ratio of the solution density in the most preferred area
(the darkest slice in the second quadrant) to that in the 1-$\sigma$ $\delta_{CKM}$ range is about 7 to 1.

As far as neutrino parameters are concerned, the solar mixing angle shows no
longer the tight lower bound present in the CP-conserving case,
namely $\sin^{2}2\theta_{12}|_{CP=0}>0.85$ (for
$\tan\beta =10$)~\cite{Bertolini:2004eq}. On the other hand, the
experimental improvement on the allowed values for the solar mixing
sharpens a tension with the strange quark mass.

\begin{figure}[h]
\caption{\label{sol1} A density plot of
$\sin^{2}2\theta_{12}$ is shown as a function of $\sin^2 2\theta_{23}$ in
the minimal renormalizable SUSY $SO(10)$ model with complex Yukawa couplings
for $\tan\beta = 10$.
The solid contour encloses the experimentally allowed region at the $90\%$ C.L.
The dark area corresponds to the solutions that are
consistent with all neutrino data.
}
\epsfig{file=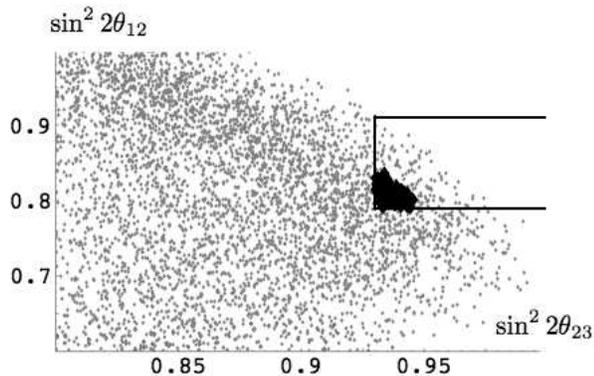 ,width=8cm}
\end{figure}

\begin{figure}[h]
\caption{\label{dms} A density plot of $|\Delta m_{\odot}^{2}/\Delta m^{2}_{A}|$ is shown as a
function of $\sin^2 2\theta_{23}$ in the minimal renormalizable
SUSY $SO(10)$ model with complex Yukawa couplings for $\tan\beta = 10$.
The solid contour encloses the experimentally
allowed region at the $90\%$ C.L. The dark area corresponds to the solutions that are
consistent with all neutrino data.
}
\epsfig{file=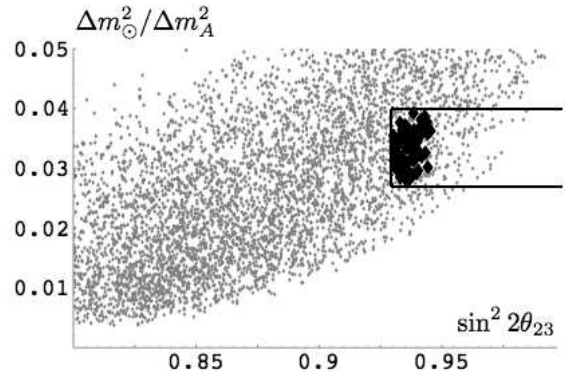 ,width=8cm}
\end{figure}

Fig.~\ref{sol1} shows the area in the solar and
atmospheric mixing angles that is consistent with the ratio of neutrino
mass square differences.
While the latter (Fig.~\ref{dms}) bounds
this area from above, we find that the 90\% C.L. lower bound on the solar mixing
requires a strange quark mass above 30 MeV at the GUT scale, that
corresponds to $m_s(2\ {\rm GeV}) > 140$ MeV (the solutions in the allowed region
span a GUT scale strange quark mass in the $30-34$ MeV range).
Should the experimental value for the solar mixing angle settle above
the present central value, it would represent a serious shortcoming of the
minimal $SO(10)$ framework. The same conclusion applies to a maximal atmospheric mixing.

\begin{figure}[h]
\caption{\label{Ue31} A density plot of $|U_{e3}|$ is shown as a
function of $\sin^2 2\theta_{23}$ in the minimal renormalizable
SUSY $SO(10)$ model with complex Yukawa couplings for $\tan\beta = 10$.
The solid contour encloses the experimentally
allowed region at the $90\%$ C.L. The dark area corresponds to the solutions that are
consistent with all neutrino data.
}
\epsfig{file=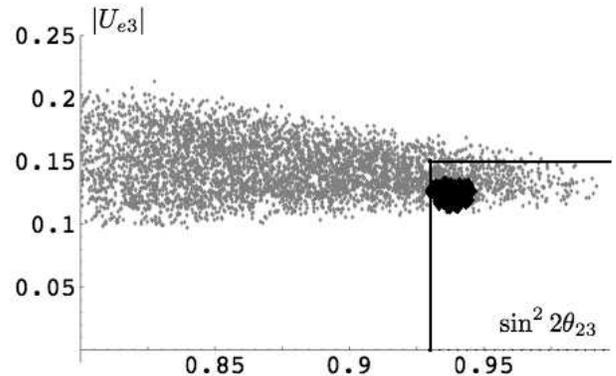 ,width=8cm}
\end{figure}

The lower bound for the
$|U_{e3}|$ parameter is relaxed compared to the CP conserving case,
although not as dramatically as for the solar angle.
As Fig. \ref{Ue31} shows, for $\tan\beta= 10$
the constraint $|U_{e3}|\geq  0.15$ found in the
CP conserving setting (see the discussion in ref.~\cite{Bertolini:2004eq})
is lowered to about $|U_{e3}|\geq  0.1$.
In the case of $\tan\beta= 55$ we find at the level of accuracy of our numerical analysis 
consistent solutions at the 95\% C.L. in the neutrino data.
On the other hand, the lower bound on $|U_{e3}|$ remains unaffected.
The persistence of such a non-vanishing lower bound is a clear signature of the
tight correlation between lepton and quark Yukawa couplings in this framework,
that makes $|U_{e3}|\simeq O(\lambda)$~\cite{Mohapatra2}.

In order to give an explicit numerical example we may consider the following (GUT scale) values
for the quark masses~\cite{Das:2000uk}
$$
\begin{array}{lll}
m_{u}\sim 0.57\, {\rm MeV}  & & m_{d}\sim 0.73\, {\rm MeV} \\
m_{c}\sim 235.7\, {\rm MeV}  & & m_{s}\sim 31.3\, {\rm MeV}  \\
m_{t}\sim 90.0\, {\rm GeV}  & & m_{b}\sim 1.19\, {\rm GeV}
\end{array}
$$
together with the CKM parameters
$$
\begin{array}{lll}
\sin\phi_{12}\sim 0.2253 & &\sin\phi_{23}\sim 0.0331  \\
\sin\phi_{13}\sim 0.0035 & &\delta_{CKM}\sim 75^{0}\  .
\end{array}
$$
For
$$
\begin{array}{lll}
\alpha_{1}\sim 144^{0}& & \beta_{1}\sim 216.8^{0} \\
\alpha_{2}\sim 142^{0} & &\beta_{2}\sim 224.6^{0} \\
\alpha_{3}\sim 1.2^{0}& & \omega \sim - 0.2^{0}\\
|r|\sim 0.748 & & |k'|\sim 0.256
\end{array}
$$
one obtains the following charged lepton
mass matrix (normalized to the $\tau$ mass):
\bea
&  100\, k' \tilde{M_{l}'}\ \approx \ |k'| \ \times & \nn \\
&  \left(
\begin{array}{ccc}
-0.146+ 0.004 i & -0.134 + 0.016 i & 0.35 + 2.862 i \\
-0.134 + 0.016 i & -6.985 - 0.197 i & 5.321  - 11.506 i \\
0.35 + 2.862 i &  5.321  - 11.506 i  & 97.846+8.567 i \\
\end{array}
\right) & \nn \eea
The corresponding (GUT-scale) charged lepton masses are all within their
90\% C.L. ranges: $m_{e}=0.3585$
MeV, $m_{\mu}=75.62$ MeV and $m_{\tau}=1 294.0$ MeV.
The neutrino mass matrix
is then given by
\bea
&  100\, \tilde{M_{\nu}'}\propto & \nn \label{neutrinomatrix} \\
&  \left(
\begin{array}{ccc}
-0.203+0.004  i & -0.134+0.016 i & -0.339 +2.863  i \\
-0.134+0.016 i & -9.404-0.224  i & 5.366-11.485 i \\
-0.339 +2.863 i  &  5.366-11.485 i & 5.855+8.952 i
\end{array}
\right)& \nn \eea
Keeping into account that
the absolute mass scale is set by the VEV of the LH triplet in
$\overline{126}_H$,
one finds neutrino mass ratios and mixings
($\sin^{2} 2\theta_{12}=0.82$,
$\sin^{2} 2\theta_{23}=0.93$, $|U_{e3}|=0.11$,
${\Delta m^{2}_{\odot}}/{\Delta m^{2}_{A}}=0.027$)
that are within the present 90\% C.L. experimental data.

The comparison with the data in Table~\ref{nudata} must account for the
running of the parameters from the GUT scale to the weak scale.
For normal hierarchy with $m_{1}^2/m_{2}^2 \ll 1$ (which is generally the case in the
setup here considered) the effects of running of the neutrino mixings and the $\Delta m^{2}$ ratios
down to the weak scale are mild \cite{Antusch:2003kp}.
In particular, $|U_{e3}|$ is very stable, with corrections below 1\% for the whole range of
$\tan\beta$ considered.
The variation of the atmospheric angle is small as well, remaining below one percent
for $\tan\beta=10$ and at the percent level for $\tan\beta=55$.
The largest corrections appear for the solar angle at large $\tan\beta$:
several percents for $\tan\beta = 55$.
Both solar and atmospheric mixing angles grow when approaching the weak scale.

As far as the leptonic CP phases are concerned, the Dirac phase $\delta_{PMNS}$ turns out to be
generally small ($< 15^o$), while the two neutrino Majorana phases $\varphi_{1,2}$ show an
approximate $180^o$ correlation.
In the example reported one finds
$\delta_{PMNS}= 4^o$, $\varphi_{1}= 10^o$, $\varphi_{2}= 191^o$.

\begin{table}
\caption{\label{nudata} Presently allowed values (90\% C.L.) for
neutrino mixing and mass parameters~\cite{Strumia:2005tc}.}
\begin{tabular}{rcl}
\hline
\hline
\null\\[-2.5ex]
$0.79 \lesssim$  & $\sin^{2}2\theta_{12}$ & $\lesssim 0.91 $ \\
$0.93 \lesssim$  & $\sin^{2}2\theta_{23}$ &\\
$|\sin \theta_{13}|$ & $\lesssim 0.15$ &\\[2ex]
\hline
\null\\[-2ex]
$2.7\times 10^{-2}\lesssim$ & $\frac{\Delta m^{2}_{\odot}}{\Delta m^{2}_{A}}$ &
$\lesssim 4.0 \times 10^{-2}$\\[2ex]
\hline
\hline
\end{tabular}
\end{table}

We conclude that the minimal SUSY $SO(10)$ GUT, when complex Yukawa couplings are
taken in their generality, is not ruled out by present data on the quark and leptons textures.
On the other hand, due to a raising tension with the strange quark mass
a large solar mixing and a maximal atmospheric angle can hardly be accomodated.
The non-vanishing lower bound for $|U_{e3}|$ is a robust prediction of the model that falls
within the reach of the planned long-baseline neutrino experiments
(a discussion on proton decay and lepton flavor violation processes
within this framework is presented in refs.~\cite{Goh:2003nv,Mohapatra3}).

\section{The model with quasi-decoupled 120-dimensional Higgs representation \label{modelwith120}}
In the second part of this paper we consider the case of complex
Yukawa couplings in a simple extension of the minimal renormalizable
$SO(10)$ proposed in ref.~\cite{Bertolini:2004eq}. The minimal
setting is enlarged to include a $120_H$ chiral super multiplet to
which the $16_F\times 16_F$ matter bilinear couples.
At variance with ref.~\cite{Dutta:2004hp} no a-priori restrictions are imposed
on the form of the Yukawa couplings. On the other hand,
in ref.~\cite{Bertolini:2004eq} the $120_H$
contributions to the fermion masses are assumed to be
two to three orders of magnitude smaller than those induced by $10_H$
and $\overline{126}_H$.
This can be seen as a consequence of a partial decoupling of the $120_H$ multiplet
and/or a small Yukawa coupling. In both cases the setup remains stable under
quantum corrections and it is (technically) natural.
It is then shown that, due to the different
symmetry property of the $120_H$ Yukawa coupling, the related
contributions to the mass matrices affect the neutrino mixing angle predictions in such a
way to improve substantially the agreement with the data, even at
the 1-$\sigma$ level. The model does not lose its predictivity and
in particular the sharp prediction for the $U_{e3}$ mixing angle
remains.

In this paper we discuss how the outcomes of the non-minimal setting
change by switching on CP violation in the Yukawa sector. For a
detailed discussion of the extended setup we refer the reader to
ref.~\cite{Bertolini:2004eq}. We comment here on the features that
differ from the real Yukawa case.

Once the left-handed doublets contained in $10_H$,
$\overline{126}_H$ and $120_H$ acquire a vacuum expectation value
(VEV), the contributions to the quark and lepton mass matrices are
generalized to \bea
\label{relations} M_u &=& Y_{10}v_u^{10}+ Y_{126}v_u^{126}+ Y_{120} v_u^{120} ~,\nn \\
M_d &=& Y_{10}v_d^{10}+ Y_{126}v_d^{126}+ Y_{120} v_d^{120} ~, \\
M_l &=& Y_{10}v_d^{10}-3 Y_{126}v_d^{126}+ Y_{120} v_l^{120} ~,\nn\\
M_\nu &=& Y_{126}v_T^{126} ~\nn. \eea
While $Y_{10}$ and
$Y_{126}$ are complex symmetric matrices, $Y_{120}$ is a complex
antisymmetric matrix. As discussed in ref.~\cite{Bertolini:2004eq},
$v^{120}_{x}$  are taken to
be suppressed with respect to $v^{10}_{x}$ and $v^{126}_{x}$ by two
to three orders of magnitude (alternatively the Yukawa coupling should
provide the required suppression).
In such a case the
$Y_{120}$-proportional terms in (\ref{relations}) can be treated as
perturbations of the minimal model results. This allows us to
maintain the successful leading order features of
the model while treating the multi-parameter
problem via a perturbative approach.

The generalized sum rules for the charged lepton and neutrino mass
matrices then read \bea k \tilde{M_l} & = & \tilde{M_u} + r
\tilde{M_d}+Y_{120}(k \varepsilon_l-\varepsilon_u-r \varepsilon_d)
~, \label{simplified} \\
\tilde{M}_\nu &\propto &\tilde{M}_l - Y_{120}\varepsilon_l
-\frac{m_b}{m_\tau} \left( \tilde{M}_d-Y_{120}\varepsilon_d\right)
\nn \eea where
$$
\varepsilon_u\equiv \frac{v_u^{120}}{m_t} ~,\qquad
\varepsilon_d\equiv \frac{v_d^{120}}{m_b} ~,\qquad
\varepsilon_l\equiv \frac{v_l^{120}}{m_\tau} ~.
$$
As before, one can rotate away the phase of $r$ and diagonalize all
the quark mass matrices by means of {\em biunitary} transformations
$\tilde{M}_{x}={V_{x}^{R}}\tilde{D}_{x}{V_{x}^{L}}^{T}$. The sum
rules in \eq{simplified} can be then recast as follows: \bea
k' {V^R_d}^\dagger\tilde{M}_l{V^L_d}^{*} & =
&{V^R_d}^\dagger{V^R_u}\tilde{D}_u
V_{CKM}^{0}- |r| \tilde{D}_d \nn \\
\label{chargedleptonformula} & +& Y_{120}' (k' \varepsilon_l-{\rm
e}^{-i\phi_{r}}\varepsilon_u+|r|
\varepsilon_d) \\
k'{V^R_d}^\dagger{M}_\nu{V^L_d}^{*} &\propto & k'
{V^R_d}^\dagger\tilde{M}_l{V^L_d}^{*} - Y'_{120}k'\varepsilon_l
\nn \\
&-&|k'| {\rm e}^{i\omega}\left|\frac{m_b}{m_\tau}\right| \left(
\tilde{D}_d-Y'_{120}\varepsilon_d\right)\ , \eea where $k'\equiv k
{\rm e}^{-i\phi_{r}}$, $Y_{120}'\equiv {V^R_d}^\dagger Y_{120}
{V^L_d}^{*}$ and ${V_{u}^{L}}^{T}{V_{d}^{L}}^{*}\equiv
V_{CKM}^{0}=P_{u}V_{CKM}P_{d}$.
 Since the antisymmetric components in
\eq{relations} are very small, the right-handed quark
mixing matrix $W\equiv{V^R_u}^T{V^R_d}^{*}$ can be estimated
perturbatively~\cite{Bertolini:2004eq}. In the complex case the
relevant equation reads ($x=u,d$): \bea
W=&V_{CKM}^{0}+2\left(-|\varepsilon_{u}|{Z'_{u}}^{*}V_{CKM}^{0}+|\varepsilon_{d}|V_{CKM}^{0}{Z'_{d}}^{*}\right)\nn
\eea
where $Z_{x}'$ are antihermitean matrices ($M^{T}=-M^{*}$) obeying
\bea Z_{u}' \tilde{D}_{u}+\tilde{D}_{u}{Z_{u}' }^{*}&=&{\rm
e}^{i\phi_{u}}{V_{CKM}^{0*}}Y_{120}'{V_{CKM}^{0\dagger}}\equiv
A_{u}\nn
\\
\label{Zprimes} Z_{d}' \tilde{D}_{d}+\tilde{D}_{d}{Z_{d}'
}^{*}&=&{\rm e}^{i\phi_{d}}Y_{120}' \equiv A_{d}
\eea
and $\varepsilon_{u,d}={\rm e}^{i\phi_{u,d}}|\varepsilon_{u,d}|$.
Eqs. (\ref{Zprimes}) are then solved by
\bea
{\rm Re}(Z'_{x})_{ij}=\frac{{\rm Re}(A_{x})_{ij}}{(\tilde{D}_{x})_{ii}+(\tilde{D}_{x})_{jj}} \nn \\
{\rm Im}(Z'_{x})_{ij}=\frac{-{\rm
Im}(A_{x})_{ij}}{(\tilde{D}_{x})_{ii}-(\tilde{D}_{x})_{jj}} \eea
Having set all relevant notation one can perform a numerical fit for
a given set of the additional parameters $Y_{120}'$,
$\varepsilon_l$, $|\varepsilon_{u,d}|$ and $\phi_{u,d}$. We will
follow closely the numerical analysis of
ref.~\cite{Bertolini:2004eq}.

\subsection{Electron mass formula and screening of the CKM CP violating phase}
Using the hierarchical structure of the
RHS of \eq{chargedleptonformula} one can again expand the magnitude
of the normalized electron mass in powers of $\lambda$:
\be
\label{mewith120} |k'\,\tilde{m}_{e}|{\rm e}^{i \phi}= {T_{\rm
MM}}+{\Delta T_{120}}
\ee
Where the symbol ${T_{\rm MM}}$ stands for the
minimal model contribution (the RHS of \eq{meminimal}). The
correction coming from the additional terms in eq.
(\ref{chargedleptonformula}) reads
\bea {\Delta
T_{120}}=-\frac{|r|}{F_{s}}{\rm
e}^{i\beta_{1}}F_{\varepsilon_{d}}^{2}[(Y_{120}')_{12}]^{2}\lambda^{5}
+{\cal O}(\lambda^{6}).
\eea
where
$F_{s}\equiv \frac{m_{s}}{m_{b}}/\lambda^{3}$
and
$F_{\varepsilon_{d}}\equiv \varepsilon_{d}/\lambda^{4}$
are ${\cal O}(1)$ form factors~\cite{Bertolini:2004eq}.

Notice that the  ${\cal O}(\lambda^{5})$ term of  ${\Delta T_{120}}$
is in general larger than the $\Lambda^2$ term on the RHS of
\eq{meminimal} and thus the partial cancellation of the leading
${\cal O}(\lambda^{4})$ in \eq{mewith120} is more easily achieved.
As a consequence, one may expect the CKM phase not to be biased towards
unphysical values, as it happens in the minimal setup.

By inspection of the ${\cal O}(\lambda^{5})$ term in $\Delta
T_{120}$ and the leading ${\cal O}(\lambda^{4})$ term in
(\ref{meminimal}), a possible way to make these two terms
interfere destructively is by taking purely imaginary entries of the
Yukawa matrix $Y_{120}'$, while assuming no spontaneous CP
violation.
This particular form of the coupling is actually obtained
in ref.~\cite{Dutta:2004hp} via an additional parity symmetry that forces all
Yukawa interactions to be hermitian.
We are now ready to discuss the numerical results.
\subsection{Numerical results}
In analogy with the discussion of the CP conserving case we present a sample
of the numerical outcomes for a given set of the $120_{H}$ parameters.
According to the previous discussion we take
$$
Y_{120}'= i \left(
\begin{array}{ccc}
0 & 1 & -1 \\
. & 0 &  1 \\
. & . & 0
\end{array}
\right),
$$
$$
|\varepsilon_{d}|=10^{-3},\,\, |\varepsilon_{u}|=10^{-4},
\,\,\phi_{u,d}=0\ ,
$$
and $\varepsilon_{l}=0$ for simplicity.

\begin{figure}[h]
\caption{\label{density120} The relative densities of the charged lepton
mass solutions are shown  as a function of the CKM phase $\delta$
in the extended $120_H$ model, for the setup described in the text and
90\% C.L. input data from Table \ref{GUTdata}.
}
\epsfig{file=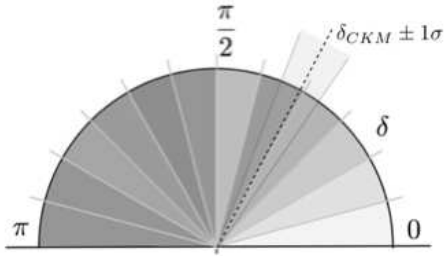 ,width=6cm}
\end{figure}

\begin{figure}[h]
\caption{\label{Ue3120} A typical density plot of $|U_{e3}|$ is shown as a
function of $\sin^2 2\theta_{23}$
in the renormalizable SUSY $SO(10)$ model with complex Yukawa couplings
and an additional $120_{H}$ Yukawa term.
The solid contour encloses the experimentally allowed region at the $90\%$ C.L.
}
\epsfig{file=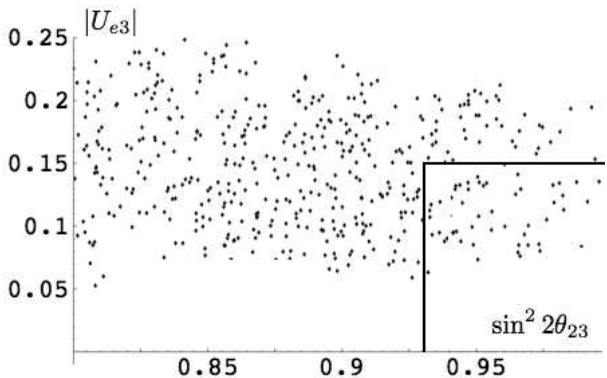 ,width=8cm}
\end{figure}

In Fig. \ref{density120} we display the relative densities of the charged lepton
mass solutions as a function of the standard CKM phase.
 As expected, the tension driving the CKM phase to
the second or third quadrant is almost completely
screened by the new $120_H$ induced terms, making
the physical $\delta_{CKM}$ a natural outcome of the
numerical scan. Correspondingly, we find that the presently allowed ranges
for the solar and atmospheric mixing can be fully covered by tuning
the additional contributions. The analogues of Figs. \ref{sol1} and \ref{dms} do not add
more information and we omit them here.

It is however worth emphasizing that in spite of the
additional freedom in the parameter space a non-vanishing lower bound remains for the
$|U_{e3}|$ mixing angle, as in the CP conserving case~\cite{Bertolini:2004eq}.
As shown in Fig.~\ref{Ue3120}, obtained
for 90\% C.L. input data and for the $120_H$ setup
discussed above, we find $|U_{e3}|\gtrsim 0.05$.
The bound remains stable
for $\tan\beta$ in the 10 to 55 range.

\section{Conclusions}
We have re-examined the role of CP phases within the minimal
renormalizable SUSY $SO(10)$ grand unified model, paying particular attention to the fit of
the charged lepton masses and the neutrino data.
We have shown, against some prejudice present in previous studies,
that the observed CP violation in the quark
sector can be fully accounted for by the standard CKM phase,
without a severe fine tuning on the charged lepton data.
While the neutrino $\Delta m^2$ ratio covers all of the present 90\% C.L. range,
both solar and atmospheric angles are obtained in their lower ranges.
Finally we emphasize the sharp prediction for $|U_{e3}|$ as a characteristic and
robust signature of the model.

As far as the renormalizable extension introduced in ref.~\cite{Bertolini:2004eq}
is concerned, we have found that
the fine tuning in the electron mass is dramatically reduced by the presence of small
contributions coming from the additional Yukawa term. The $120_H$ induced
corrections may extend the predictions for the solar and atmospheric
neutrino mixings to cover the present experimental ranges.
In spite of the additional parameters a robust $|U_{e3}|$ lower bound remains,
that characterizes the mass sum rules of the minimal renormalizable  $SO(10)$ setup.

\vspace*{-1ex}
\section*{Acknowledgments}
\vspace*{-2ex}
We thank Borut Bajc, Bhaskar Dutta, Michele Frigerio, Goran Senjanovi\'c  and Francesco Vissani
for helpful discussions and comments on the manuscript.

\end{document}